\documentclass[twocolumns,a4paper]{iopart}
\usepackage{graphicx}
\usepackage{dcolumn}
\usepackage{bm}

\begin{document}

\title[Magnetism in V, Cr and Mn chains on Cu(001)]{Magnetic anisotropy and spin-spiral wave in V, Cr and Mn atomic chains on Cu(001) surface: First principles calculations}

\author{J. C. Tung$^1$, Y. K. Wang$^1$ and G. Y. Guo$^{2,3}$\footnote{E-mail: gyguo@phys.ntu.edu.tw} }
\address{$^1$Center for General Education and Department of Physics, National Taiwan Normal University, Taipei 106, Taiwan\\
$^2$Graduate Institute of Applied Physics, National Chengchi University, Taipei 116, Taiwan\\
$^3$Department of Physics and Center for Theoretical Sciences, National Taiwan University, Taipei 106, Taiwan}

\begin{abstract}

Recent {\it ab intio} studies of the magnetic properties of all 3$d$
transition metal (TM) freestanding atomic chains predicted that these nanowires could
have a giant magnetic anisotropy energy (MAE) and might support a spin-spiral structure, thereby suggesting
that these nanowires would have technological applications in, e.g., high density magnetic data storages.
In order to investigate how the substrates may affect the magnetic properties of the
nanowires, here we systematically study the V, Cr and Mn linear atomic chains on the Cu(001) surface based on the
density functional theory with the generalized gradient approximation. We find that V, Cr, and Mn linear chains on
the Cu(001) surface still have a stable or metastable ferromagnetic state. However, the ferromagnetic state is
unstable against formation of a noncollinear spin-spiral structure in the Mn linear chains and also
the V linear chain on the atop sites on the Cu(001) surface, due to the frustrated magnetic interactions
in these systems. Nonetheless, the presence of the Cu(001) substrate does destabilize the spin-spiral state
already present in the freestanding V linear chain and stabilizes the ferromagnetic state in the V linear chain
on the hollow sites on Cu(001). When spin-orbit coupling (SOC) is included, the spin magnetic moments
remain almost unchanged, due to the weakness of SOC in 3$d$ TM chains. Furthermore,
both the orbital magnetic moments and MAEs for the V, Cr and Mn are small, in comparison with
both the corresponding freestanding nanowires and also the Fe, Co and Ni linear chains on the Cu (001) surface.
\end{abstract}


\maketitle

\section{Introduction}
Nanostructured magnetic materials have recently received enormous attention because of their fascinating
physical properties and potential applications.  Finite free-standing gold atomic chains in the break-junction
experiments were first reported in 1998\cite{Ohnishi,Yanson}. However, these freestanding atomic chains are unstable and
transient. Physically stable magnetic nanowires deposited on metallic substrates are one of the most important magnetic
nanostructures, and many techniques have been used to prepare and study them.
In particular, Gambardella {\it et al} \cite{Gambardella1,Gambardella2} succeeded in preparing
a high density of parallel atomic chains along steps by growing Co on a high-purity Pt (997) vicinal surface in a narrow
temperature range of 10-20 K. The magnetism of the Co wires was also investigated by the x-ray magnetic circular
dichroism.\cite{Gambardella2} Recently, the Fe double chains deposited on the Ir(001) surface
were prepared\cite{Ham03}, and their structures were investigated by both the scanning tunneling microscopy measurements 
\cite{Ham03} and the theoretical calculations\cite{Mazzarello09}.

More recently, we have performed systematic {\it ab initio} studies of the magnetic anisotropy\cite{Tung} and spin-spiral
wave in all 3$d$ transition metal (TM) freestanding linear chains. Interestingly, we found that
the Fe and Ni freestanding linear chains have a gigantic magnetic anisotropy energy (MAE)\cite{Tung}, and that
the magnetic couplings in the V, Mn and Fe linear chains are frustrated, resulting in the formation of stable
spin-spiral structures\cite{Tung3}. Saubanere {\it et al} also found a stable spiral magnetic order in
the freestanding V atomic chain in their recent {\it ab initio} theoretical study of TM nanowires\cite{Sau10}.
Experimentally, copper and tungsten are excellent substrates for growth of Fe thin
films\cite{Hauschild, Tian} because of the small lattice constant mismatch between Fe and Cu (3.61\AA) as well as
W (3.61\AA). This stimulated, e.g., a recent theoretical study of the magnetic order and exchange interactions of
the 3$d$ TM monoatomic chains on the (110) surface of Cu, Pd, Ag, and NiAl\cite{Mokrousov07}. 
Here in this paper, we perform first principles calculations for the magnetic moments, 
magnetic anisotropy energies and spin-spiral wave energies of the V, Cr and Mn linear chains deposited on the Cu (001) 
surface, in order to study how the presence of the substrate would modify the magnetic properties of the V, Cr and Mn nanowires.

\section{Theory and Computational Method}

\begin{figure}
\includegraphics[width=10cm]{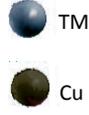}
\caption{(color online) Theoretical atomic structure used for the present slab supercell modeling of the 3$d$ 
transition metal (TM) V, Cr and Mn linear chains on the hollow sites of the Cu (001) surface. The rectangular 
box frame represents the primitive cell of the three-dimensional periodic slab supercell structure.}
\label{Cstru}
\end{figure}

The present calculations are based on density functional theory with the generalized gradient approximation (GGA)\cite{Perdew91}.
The accurate frozen-core full-potential projector augmented-wave (PAW)
method,~\cite{blo94} as implemented in the Vienna {\it ab initio} simulation package (VASP) \cite{vasp1,vasp2}, is used.
A large plane-wave cutoff energy of 350 eV is used for all the systems considered. The V, Cr and Mn linear atomic chains
along the {\it x} direction on the Cu (001) surface are modeled by a nanowire attached to both sides of a seven-layer-thick
Cu (001) slab, as plotted in Fig. \ref{Cstru}. The transition metal (TM) atoms on the nanowires are placed either on the top
of surface Cu atoms [denoted here as the atop (A) site] or at the hollow position on the Cu surface [called here as hollow
(H) site]. The two-dimensional unit cell is chosen to be of $p$(4$\times$1) structure. The nearest in-plane (out of plane)
wire-wire distance is larger than 10\AA$ $ (11\AA) which is sufficiently wide to decouple the neighboring wires. The theoretical
lattice constant (3.60~\AA) of bulk copper, which is in good agreement with experimental Cu lattice constant of 3.61~\AA,
is used as the fixed in-plane lattice constant of the Cu slab. However, the atoms are allowed to move in the surface-normal
direction, and the structural relaxations are performed using the conjugate gradient method. We focus on the nonmagnetic and
ferromagnetic states of the V, Cr and Mn linear chains on Cu (001) surface. 
The equilibrium structure is obtained when all the forces acting on the atoms are less than 0.02 eV/\AA$ $.
The $\Gamma$-centered Monkhorst-Pack scheme with a $k$-mesh of $20\times5\times 1$ in the full Brillouin zone (BZ), in
conjunction with the Fermi-Dirac-smearing method with $\sigma = 0.2$ eV, is used for the BZ integration.

We also consider the transverse spin-spiral states where all the spins rotate in a plane perpendicular to the atomic chain axis.
The generalized Bloch theorem \cite{Herring96,Sandratskii93} is used to calculate self-consistently the total energies of 
the transverse spin-spirals as a function of the spin-spiral wave vector.
Therefore, the relativistic spin-orbit coupling (SOC) is not included in these calculations.
A denser 25$\times$5$\times$1 $k$-point mesh is used for the V and Mn chains, in order to ensure that the calculated spin wave 
excitation energies (see Fig. 3a below) and also the first and
second near-neighbor exchange interaction parameters (i.e., $J_{01}$ and $J_{02}$ in Table 2 below) are converged
to within a few percents.

{\it Ab initio} calculation of the MAE is computationally very demanding because of the smallness
of MAE, and thus needs to be carefully carried
out.\cite{guo91} Here we use the total energy difference approach rather than the widely used force theorem to determine the MAE, i.e.,
the MAE is calculated as the difference between the full self-consistent total energies for the two different magnetization directions
concerned. The total energy convergence criterion is 10$^{-8}$ eV/atom. The same $k$-point mesh is used for the density of states
calculations. The MAEs calculated with a denser 32$\times$6$\times$1 $k$-point mesh hardly differ from that obtained with the
20$\times$5$\times$1 $k$-point mesh (within 0.01 meV per magnetic atom).

\section {Results and Discussion}

\subsection {Formation energy and spin magnetic moment}

\begin{table}
\caption{Calculated formation energy $E^f$, equilibrium interlayer distance $d_{eq}$, spin magnetic moment
per magnetic atom, $m_s$, and magnetization energy per magnetic atom $E^{mag}$=$E^{FM}$-$E^{NM}$ of the 3$d$ TM nanowires on
both the hollow (H) and atop (A) sites of the Cu (001) surface. Here superscripts $FM$ and $NM$ denote the ferromagnetic and
nonmagnetic states, respectively.}
\begin{tabular*}{\textwidth}{@{}l*{15}{@{\extracolsep{0pt plus 12pt}}l}}
\mr
      & site & $E^f$& $d^{NM}_{eq}$ & $d^{FM}_{eq}$ & $m_s$ &$E^{mag}$\\
      &      &  (eV/u.c.) & (\AA)   &(\AA)  & ($\mu$$_B$)& (eV)   \\\hline
   V  & H & -1.29 & 1.64  & 1.70& 2.11& -0.559\\
      & A & -0.01 & 1.83  & 2.03& 3.30& -0.727\\
   Cr & H & -1.56 & 1.56  & 1.80& 4.23& -0.769\\
      & A & -0.86 & 1.70  & 2.18& 4.60& -1.222\\
   Mn & H & -0.52 & 1.49  & 1.73& 4.14& -0.613\\
      & A & -0.25 & 2.25  & 2.37& 4.53& -0.458\\
   Fe$^{\rm a}$ & H &  -2.14& 1.51 & 1.64& 3.07& -0.337\\
      & A &  -0.31&  1.79 & 2.31& 3.29& -0.444\\
   Co$^{\rm a}$& H &  -2.15& 1.50 & 1.57& 1.79& -0.130\\
      & A &  -1.19&  1.80 & 2.27& 1.99& -0.205\\
   Ni$^{\rm a}$& H &  -2.26&  1.55& 1.55& 0.00&  0.000\\
      & A &  -1.32&  1.96 & 2.26& 0.65& -0.032\\
\br
\end{tabular*}
\begin{itemize}
\item[] $^{\rm a}$ Theoretical calculations (Ref. \cite{Tung2})
\end{itemize}
\label{table1}
\end{table}

To see how the Cu substrate interacts with the V, Cr and Mn chains, we first study the chain
formation energy. The chain formation energy $E^f$ describes the difference in the total energy of
the combined system of the Cu substrate and a TM chain before and after the TM chain
is deposited on the substrate. As in Ref. \cite{Tung3}, we introduce the chain formation energy
as $E^f =\frac{1}{2}({E_t-nE^{bulk}_{Cu}-mE^{chain}_{TM}})$ where $E_t$ is the total energy of the system
in the ferromagnetic (FM) state, $E^{bulk}_{Cu}$ is the
total energy of the bulk Cu, and $E^{chain}_{TM}$ is the total energy of the freestanding
transition metal nanowires in the FM state. And $n$ and $m$ are the numbers of the Cu and TM atoms in the system,
respectively. It should be noted that there is no universal definition of the chain formation energy.
The calculated formation energies of all the V, Cr and Mn nanowires are listed in Table \ref{table1}.
The formation energies for the Fe, Co and Ni linear chains on the Cu(001) surface
reported recently\cite{Tung2} are also listed in Table \ref{table1} for comparison.
It is clear from Table \ref{table1} that it is more energetically favorable when the TM
atoms are placed on the hollow sites, as might be expected because the hollow site has a higher
coordination number. Among all the 3$d$ TM nanowires, Ni nanowire is most energetically favorable on both
the hollow and atop sites. Generally, the V, Cr and Mn chains are less energetically favorable than the corresponding
Fe, Co and Ni chains (Table \ref{table1}). Interestingly, the formation energy of the V chain
on the atop site is almost zero, thereby suggesting that this chain would not be stable.

Table \ref{table1} shows that the interlayer distance between the V, Cr, and Mn nanowires
on the hollow (atop) site and the Cu substrate is 1.70 (2.03), 1.80 (2.18), and 1.73 (2.37)~\AA,
respectively. The ideal interlayer distance for Cu substrate is 1.81~\AA.
Therefore, the copper substrate seems to pull (push) the TM nanowires when deposited on
the hollow (atop) site.
In all the cases considered here, the equilibrium interlayer distance is larger in
the FM state than in the nonmagnetic (NM) state. This is due to the larger kinetic energy in a
magnetic state which make magnetic materials softer and larger in size (see Table \ref{table1}).
Table \ref{table1} also shows that when the interlayer distance changes, the spin magnetic moment
changes as well. In general, for all the 3$d$ TM nanowires considered, an increase in interlayer
distance will result in an increase in the spin moment. This is because the stronger overlap between
TM 3$d$ orbitals and the Cu substrate would result in a decrease in the magnetic moment.
The interatomic distance between the deposited TM atoms in all the cases considered here is 2.55 (\AA).
The calculated spin moments of the V, Cr and Mn atoms on the hollow (atop) sites
are 2.11 (3.30), 4.23 (4.60) and 4.14 (4.53) $\mu_B$/atom, respectively.
In comparison, the calculated spin moments at the same bond length (2.55 \AA) of
the freestanding V, Cr, Mn, Fe, Co and Ni nanowires reported recently\cite{Tung},
are 4.07, 5.06, 4.53, 3.30, 2.30 and 1.14 $\mu_B$/atom, respectively. Clearly,
placing the 3$d$ TM nanowires on the hollow sites significantly reduces or even quenches the
spin moments on the nanowires, whilst the spin moments are much less affected when
the nanowires are deposited on the atop sites.

\begin{figure}
\includegraphics[width=10cm]{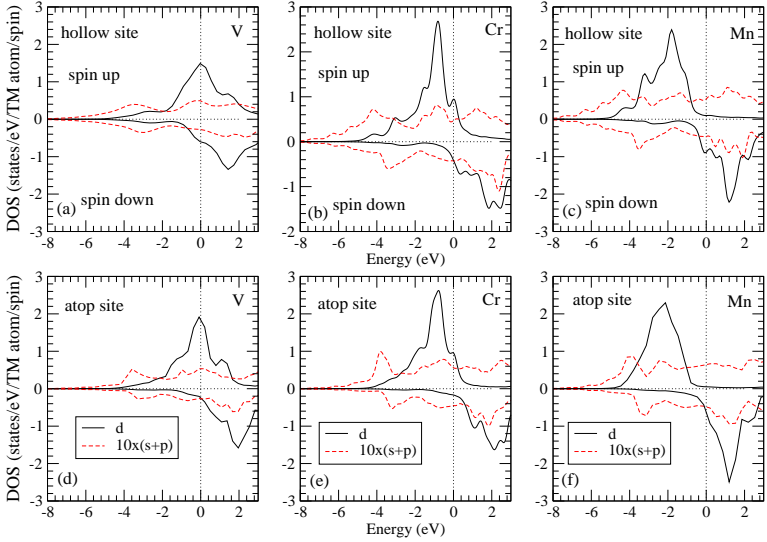}
\caption{(Color online) Spin-polarized density of states (DOS) of the V, Cr and Mn nanowires on Cu (001).
the Fermi level is at 0.0 eV. The DOS spectra for the minority spin are multiplied by $-1$, and
the $sp$-orbital decomposed DOS are scaled up by a factor of 10 for clarity.}
\label{Cdos}
\end{figure}

The obtained magnetic moments of the V, Cr and Mn nanowires can be understood in terms of
the calculated spin-polarized density of states (DOS), as displayed in Fig. \ref{Cdos}.
In Fig. \ref{Cdos}, the Fermi level is set to be zero, the DOS spectra for the minority spin
are multiplied by $-1$, and the $sp$-orbital decomposed DOS are scaled up by a factor of 10
for clarity. Clearly, the $d$-orbitals of the V, Cr and Mn nanowires on the both sites are
significantly localized due to the reduction of the coordination number whilst the $sp$-orbitals
are more dispersive. The reduction in coordination number thus induces
considerable enhancement in the spin splitting of the V, Cr and Mn 3$d$-bands.
The spin-splitting of the V, Cr and Mn 3$d$-bands for the hollow (atop)
site are 1.48 (2.01), 2.65 (2.93) and 2.98 (3.30) eV, respectively. The interlayer distance between
the TM nanowires and Cu substrate are larger on the atop site.
This indicates that the overlaps between the TM and substrates are smaller, and hence
the spin magnetic moments are larger. The splitting of the 3$d$-band is approximately
proportional to the spin moment.

\subsection {Spin-spiral wave and exchange interaction}
\begin{figure}
\includegraphics[width=10cm]{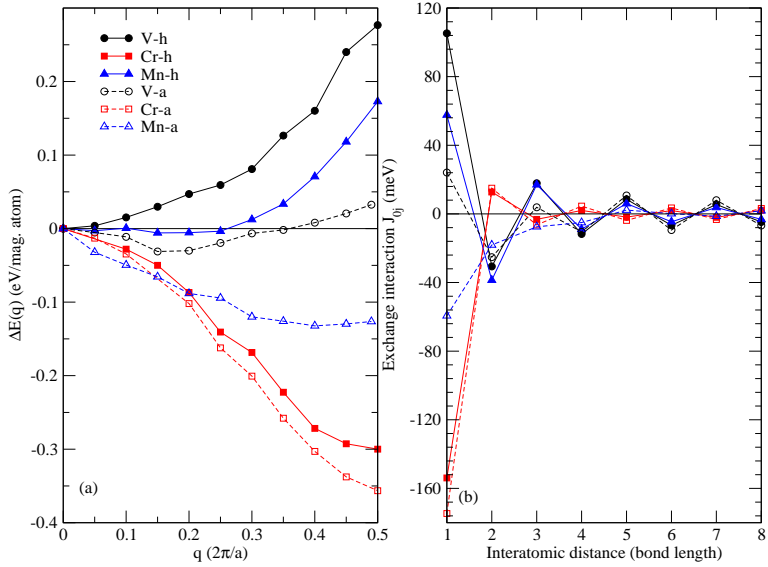}
\caption{(Color online) (a) Calculated total energies $\Delta E(q)$ [relative to the total energy of the ferromagnetic state $E(0)$,
i.e., $\Delta E(q) = E(q)-E(0)$] of the V, Cr and Mn linear chains on the Cu (001) surface as a function of spin-spiral wave vector $q$ (2$\pi$/$a$).
(b) Exchange interaction parameters $J_{0j}$ in the V, Cr and Mn linear chains on Cu(001) derived from (a) (see text).}
\label{SpiralE}
\end{figure}

In our recent study of the noncollinear magnetism in freestanding 3$d$ TM nanowires\cite{Tung3}, the V, Mn and Fe chains were found
to have a stable noncollinear spin spiral structure due to the frustrated exchange interactions in these systems whilst the Cr chain
would have the antiferromagnetic (AF) ground state. Therefore, here we perform total energy calculations for the spin spiral structure
in the V, Cr and Mn chains on Cu(001) to examine how these interesting magnetic properties would be modified in the presence of
the Cu substrate. The calculated total energies [$E$($\bf{q}$, $\theta$)] of the V, Cr and Mn linear chains on Cu(001)
are plotted as a function of the spin-spiral wave vector $q$ in Fig. \ref{SpiralE}(a). Since here only the transverse spin-spiral waves
along the chain direction (i.e., $x$-axis)
are considered, the angle between the chain axis and the magnetization direction $\theta$ = $\pi$/2,
and hence we simply write $E$($\bf{q}$, $\theta$ = $\pi$/2 ) = $E(q)$. The spin-spiral structure at $q = 0$ corresponds to the collinear
FM state, whilst the state at $q = 0.5$ (2$\pi$/$d$) corresponds to the AF state.
Fig. \ref{SpiralE}(a) shows that the FM state in all the 3$d$ TM chains considered here except the hollow-site V chain,
is unstable against the formation of a spin-spiral structure. In the Cr chains on Cu(001), the lowest energy spin-spiral state
corresponds to the AF state (i.e., $q =0.5$), being the same as in the freestanding Cr chain case\cite{Tung3}.
For comparison, we notice that in previous GGA calculations\cite{Guo00}, the FM state could not be stabilized in
bulk Cr metal, whilst the magnetization energy of the AF state is rather small ($\sim$0.016 eV/atom). 
The genuine stable spin-spiral state occurs in the atop-site V chain and hollow-site (atop-site) Mn chain on Cu(001) with $q =0.15$ and
$q =0.15$ $(0.40)$, respectively. We note that the freestanding V and Mn linear chains at the equilibrium bondlength has
the stable spin-spiral state with $q =0.25$ and $q =0.33$, respectively.\cite{Tung3}
Therefore, it appears that depositing the V chain on the hollow sites would destabilize the noncollinear spin-spiral
state and stabilize the ferromagnetic state.

The energy of a spin-wave excitation (i.e., the magnon dispersion relation) $\varepsilon(q)$ [or $\hbar\omega(q)$] can be related to
the calculated total energy of the spin-spiral state as\cite{Rosengaard97,Niu99,Tung3}
\begin{equation}
\varepsilon(q)=\frac{4\mu_B}{m_{s0}}[E(q)-E(0)]
\end{equation}
where $m_{s0}$ is the spin magnetic moment per site at $q$ = 0.
In the range of small $q$, $\varepsilon$($q$) = $D$$q$$^2$, where the spin-wave stiffness constant $D$ relates the spin-wave energy $\varepsilon$($q$)
to the wave vector $q$ in the long wavelength limit.
The spin-wave stiffness constant $D$ of an atomic chain can be estimated by fitting an even order polynomial to the corresponding spin-wave
spectrum shown in Fig. \ref{SpiralE}(a). The spin-wave stiffness constant $D$ obtained in this way for the V, Cr and Mn linear chains on Cu(001)
are listed in Table \ref{table2}. A negative value of $D$ means that the FM state is not stable against a spin-spiral wave excitation.
As mentioned above, the FM state in all the atop-site TM chains and also in the hollow-site Cr chain
is unstable and hence the $D$ for these nanowires is negative (Table \ref{table2}). Interestingly, the $D$ for the hollow-site Mn chain
is nearly zero and this is because the $E(q)$ curve is very flat in the range of $0 < q < 0.2$ [Fig. \ref{SpiralE}(a)].
Furthermore, when deposited on the hollow sites on Cu(001), the spin-wave stiffness $D$ of the V chain changes from -424 meV\AA$^2$
(freestanding chain)\cite{Tung3} to 345 meV\AA$^2$.

\begin{table}
\caption{Calculated exchange interaction parameters ($J_{0j}$) (meV) between two $j^{th}$ near neigbors ($j = 1, 2, 3, 4, 5$)
and spin wave stiffness constant $D$ (meV\AA$^2$) as well as ground state spin-spiral wave vector $q$ ($2\pi/d$) ($d = 2.55$ \AA$ $
is the interatomic distance along the chain) in the V, Cr, and Mn atomic chains on Cu (001).}
\begin{tabular*}{\textwidth}{@{}l*{15}{@{\extracolsep{0pt plus 12pt}}l}}
\mr
        & $J_{01}$  &$J_{02}$ & $J_{03}$ &$J_{04}$&  $J_{05}$ &   $D$ & $q$ \\ \hline
        &           &         & H-site   &        &           &       & \\
V       & 105.2     & -30.6   &  17.8    & -11.6  &   8.4     &  345 & 0.00 \\
Cr      & -154.0    &  12.8   &  -3.2    &  2.0   &  -1.8     & -363 & 0.50 \\ 
Mn      &  57.6     & -38.6   &  17.0    & -9.2   &   6.0     &  $\sim$0 & $\sim$0.15 \\ 
        &           &         & A-site   &        &           &      \\
V       & 24.2     & -25.2    &   3.8    & -10.4  &   10.8    & -379 & 0.20 \\ 
Cr      & -174.6    & 15.0    &  -6.4    &  4.6   &   -3.8    & -399 & 0.50 \\ 
Mn      & -59.4     & -18.2   &  -7.6   & -5.0   &    2.2   & -683 & 0.40 \\ 
\br
\end{tabular*}
\label{table2}
\end{table}

In the frozen magnon approach, the exchange interaction parameters $J_{ij}$ between atom $i$ and atom $j$ on a TM chain
are related to the magnon excitation energy $\varepsilon$($q$) by a Fourier transformation
\begin{equation}
J_{0j}=\frac{1}{N_q}\sum_{q}e^{-i{\bf q}\cdot{\bf R}}J(q)
\end{equation}
where $N_q$ is the number of $q$ points in the Brillouin zone included in the summation.
$J(q)$ is the Fourier transformation of the exchange parameters $J_{ij}$ and is related to the magnon excitation energy $\varepsilon$($q$) by
\begin{equation}
\varepsilon(q)=-\frac{2\mu_B}{m_{s0}}J(q).
\end{equation}
Here the negligible induced magnetization on all the Cu atoms is neglected and hence only the TM atoms on the deposited chain
are considered in the summation. The calculated exchange parameters $J_{ij}$ are plotted in Fig. \ref{SpiralE}(b) and also
listed in Table \ref{table2}.
The magnetic coupling between two first near neighbors in the V chains and also the hollow-site Mn chain is ferromagnetic ($J_{01}>0$),
whilst it is antiferromagnetic ($J_{01}<0$) in the rest of the Cr and Mn chains [Fig. \ref{SpiralE}(b) and Table \ref{table2}].
Interestingly, the magnetic coupling between two second near neighbors in the V and Mn chains is antiferromagnetic ($J_{02}<0$),
whilst it is ferromagnetic ($J_{02}>0$) in the Cr chains, i.e., the second near neighbor magnetic coupling
is opposite to the first near neighbor magnetic coupling in all the nanowires considered here except the atop-site Mn chain.

\subsection {Magnetic anisotropy energy}
The relativistic SOC is essential for the orbital magnetization and magnetocrystalline anisotropy in solid, although it may
be weak in the 3$d$ TM systems. Therefore, we perform further self-consistent calculations with the SOC included
in order to study the magnetic anisotropy and also orbital magnetization of the V, Cr and Mn nanowires on Cu (001),
and the results are summarized in Table \ref{table3}. For comparison, the same results for the Fe, Co and Ni linear
chains on Cu (001) reported recently\cite{Tung2} are also listed in Table \ref{table3}. First, when the SOC is
taken into account, the spin magnetic moments for the V, Cr and Mn TM nanowires on the hollow (atop) site
are 2.11 (3.30), 4.23 (4.60), and 4.14 (4.53) ($\mu_B$/mag. atom), being almost identical to the corresponding
one obtained without SOC. This is due to the weakness of the SOC in the 3$d$ transition metals.
Nevertheless, including the SOC gives rise to orbital magnetic moments
in the 3$d$ TM nanowires and, importantly, allow us to determine the easy magnetization axis in the nanowires.
For the magnetization lies along the chain direction (i.e., the $x$-axis) and the chain is on the hollow (atop) site,
the calculated orbital moments for the V, Cr and Mn atoms are 0.01 (-0.06), -0.01 (0.01) and 0.02 (0.02)
$\mu_B$/mag. atom (see Table \ref{table3}), respectively, being small when compared with the Fe, Co and Ni chains.
For comparision, we note that the corresponding orbital moments in the freestanding
V, Cr and Mn nanowires are -0.16, 0.04, 0.02 ($\mu_B$/mag. atom), respectively.\cite{Tung}
In Ref. \cite{Tung}, it was also found that the magnetization of the freestanding 3$d$ TM
freestanding nanowires has a strong directional dependence, and that the orbital moment is larger
when the magnetization lies along the chain direction.
Table \ref{table3} shows that the directional dependence of the orbital moment is weak in
the on-hollow-site V, Cr and Mn chains but is rather significant when the V, Cr and Mn chains
are on the atop sites.


\begin{table}
\caption{Calculated spin magnetic moment per atom, $m_s$, orbital magnetic moment per magnetic atom, $m_o$, along three different
directions and magnetic anisotropy energy per magnetic atom (MAE) of the V, Cr, Mn, Fe, Co and Ni chains on Cu (001). The chain direction
is along (100), and the (010) [(001)] direction is in-plane [out of plane] but perpendicular to the chain direction.
The MAE $E^1$ is defined as $E^{100}$-$E^{001}$ and $E^2$ is $E^{100}$-$E^{010}$. The calculated MAE values
are converged to within 0.01 meV per magnetic atom with respect to the $k$-points used.}
\begin{tabular*}{\textwidth}{@{}l*{15}{@{\extracolsep{0pt plus 12pt}}l}}
\mr
        & $m_s$               &       &  $m_o$        &      &      MAE  &      \\
        & ($\mu_B$)           &       &  ($\mu_B$)    &      &  (meV)    &      \\
        &                     & 100   &    010        & 001  &   $E^1$   & $E^2$\\ \hline
        &                     &       &    H-site     &      &           &      \\
V       &           2.11      & 0.01  &    0.01       & 0.01 &   0.00    &  0.00\\
Cr      &           4.23      &-0.01  &   -0.01       &-0.01 &  -0.01    &  0.01\\ 
Mn      &           4.14      & 0.02  &    0.02       & 0.03 &  -0.01    &  0.01\\ 
Fe$^{\rm a}$      &           3.07      & 0.10  &    0.09       & 0.10 &   0.32    &  0.25\\
Co$^{\rm a}$      &           1.78      & 0.27  &    0.18       & 0.17 &  -1.17    & -1.16\\
Ni$^{\rm a}$      &           0.02      & 0.02  &    0.00       & 0.00 &  -0.53    &  0.01\\
        &                     &       &    A-site     &      &           &      \\
V       &           3.30      &-0.06  &    0.18   &-0.03 &    0.00   &  0.00\\ 
Cr      &           4.60      & 0.01  &    0.09   &-0.01 &   -0.01   &  0.01\\ 
Mn      &           4.53      & 0.02  &    0.21   & 0.03 &   -0.01   &  0.01\\ 
Fe$^{\rm a}$      &           3.28      & 0.12  &    0.11       & 0.13 &    0.38   &  0.29\\
Co$^{\rm a}$      &           1.99      & 0.19  &    0.25       & 0.12 &   -0.40   & -1.51\\
Ni$^{\rm a}$      &           0.64      & 0.13  &    0.27       & 0.11 &   -0.05   & -0.32\\
\br
\end{tabular*}
\begin{itemize}
\item[] $^{\rm a}$ Theoretical calculations (Ref. \cite{Tung2})
\end{itemize}
\label{table3}
\end{table}

The calculated MAEs of the V, Cr, Mn, Fe, Co and Ni nanowires are listed in Table \ref{table3}.
The MAEs $E^1$ and $E^2$ are defined as the energy differences $E^1$=$E^{100}$-$E^{001}$
and $E^2$=$E^{100}$-$E^{010}$,
where the $E^{100}$ is the calculated total energy when the magnetization lies along the $x$ direction.
If both the $E^1$ and $E^2$ are negative, the magnetization prefers to lie along the chain ($x$-axis) direction.
The MAEs of the V, Cr and Mn nanowires on both the hollow and atop sites are quite small.
In contrast, the MAE $E_1$ of the freestanding V, Cr and Mn chains is -0.45, -0.07
and 0.28 (meV/mag. atom), respectively, with $E^2 = 0$\cite{Tung}. In Ref. \cite{Tung2},
it was found that the Fe chain on Cu(001) has an out-of-plane anisotropy while the Co and Ni
chains on Cu (001) have an in-plane anisotropy (Table \ref{table3}). In contrast, here we find
that the Cr and Mn chains on Cu(001) have an in-plane anisotropy, however,
with the total energy changes due to the magnetization rotation from the $x$, through $y$, to  $z$ axes
being very small. Clearly, when the 3$d$ TM chains are deposited on Cu(001), the MAEs generally
become smaller due to the overlap of the wavefunctions between the TM nanowires and substrate
(Table \ref{table3}). In particular, the MAEs of the V, Cr and Mn chains become very small
when they are deposited on the Cu(001) substrate.

\section {Conclusions}
We have performed systematic {\it ab initio} GGA calculations for the V, Cr and Mn linear 
atomic chains on Cu (001) surface in order to
examine how the substrates would affect the magnetic properties of the nanowires. We found that V, Cr and Mn linear chains
on Cu (001) surface still have a stable or metastable FM state. Nonetheless, we also found that the ferromagnetic state is
unstable against formation of a noncollinear spin-spiral structure in the Mn linear chains and also
the V linear chain on the atop sites on the Cu(001) surface, due to the frustrated magnetic interactions
in these systems. The presence of the Cu(001) substrate does destabilize the spin-spiral state
already present in the freestanding V linear chain and stabilizes the ferromagnetic state in the V linear chain
on the hollow sites on Cu(001). When the spin-orbit coupling (SOC) is included for the collinear ferromagnetic state, 
the spin magnetic moments remain unchanged, due to the weakness of SOC in 3$d$ TM chains. Furthermore,
both the orbital magnetic moments and MAEs (within 0.01 meV/magnetic atom) for the V, Cr and Mn are small, in comparison with
both the corresponding freestanding nanowires and also the Fe, Co and Ni linear chains on the Cu (001) surface.

\section*{Acknowledgments}
The authors acknowledge supports from the National Science Council and the NCTS of Taiwan as well as
the computing time from the National Center for High-performance Computing of Taiwan.\\

\end{document}